# Spatial and time-dependent distribution of plasma parameters in the metal-halide arc lamp.

A. Khakhaev, L. Luizova, K. Ekimov and A. Soloviev
*Petrozavodsk State University, Russia*

The metal-halide arc lamp is an effective light source and its investigation has a long history, but even at present some authors are sure that the local thermodynamic equilibrium (LTE) model can be applied to such objects [1] whereas the others have proved experimentally that such plasma is non-equilibrium object [2-4]. However some plasma parameters (electron and high lying excited states densities as well as Hg metastable levels densities) are assumed to be in equilibrium with electron temperature and these assumptions are applied in plasma diagnostics. To verify this supposition the method of local plasma spectroscopy based on spatial and temporal distribution of spectral line profiles was developed and described here.

## 1. Experimental setup.

The experimental setup consists of a high aperture scanning diffraction spectrometer and special spatial scanning block. It allows to scan spectrum with step $\delta\lambda=0.015$ nm in the range of 300-800 nm, to scan spatial source surface radiance distribution in the perpendicular to the light source axis direction with the step $\delta x=0.004$ mm in the range of ±3cm. Light intensity is measured by a photomultiplier and amplifier with time resolution less than $10^{-3}$ s. For automation of the experiment the setup was constructed on the basis of CAMAC crate. The crate contains two modules for stepper motors control, modules for trailers detection and two analog-digital converters: the one measures light intensity and the other measures arc discharge voltage. It allows to measure the intensity in chosen phases of current period. The software for the experiment was developed in LabVIEW environment [5]. The software interacts with CAMAC by means of special program driver. In our case the software runs on the IBM-compatible computer with OS Windows 98. The software for data acquisition consists of separate modules, that are implemented as virtual instruments (VI).

Before experiment starts some devices need calibration: the optical system is calibrated on wave lengths by a source with a known spectrum, the photodetector sensitivity is calibrated by the registration of a certified temperature lamp spectrum, and the linearity of ADC, that measures light intensity, is tested. The experiment is controlled by the following VI:

*VI of spectrum scanning in the given spectral range with the given step.* It was used on preliminary stage of experiment for choosing spectral lines for further analysis and then for spectral device spread function measurement. As a light source for this purpose the spectral cadmium lamp was used. Line widths in this lamp are less than 0.001 nm. In our experimental conditions (spectral slit sizes and imaging system parameters) the spread function has width equal to 0.09 nm and in further analysis this function is accepted to be gaussian of such width.

*VI of spatial gating at given wave length.* This module is used for finding the spatial point corresponding to the center of the discharge. The following measurements are carried out for the center (x=0) and other points $x_k$ (k=1, 2...m-1), corresponding to m equidistant positions of spatial gating system.

*VI of automated spatially and spectral scanning in the given spectral and spatial intervals and time phase*. ADC samples are averaged over a given number of measurements (usually - 100). The results are saved in a file for further analysis.

*VI of data loading for subsequent analysis* (Fig. 1).

First of all this module allows to load any data acquired by other modules and to derive various information from a spectrum (e.g. maximum positions and values, line widths) by LabVIEW tools for graphic processing (scaling and cursor positioning). It allows also to join some arrays corresponding to different spectral ranges and different time moments (but spatial points must be the same) in common array for joint processing, calculate covariance matrix of this array, find its eigenvalues and eigenvectors (there is the standard LabVIEW module for this purpose "EigenValues & Vectors.vi"), and calculate the projections of each spectral point to all eigenvectors. It is the essential part of the joint processing algorithm, which will be described in details in the next section.

This module is also used for calculation of the estimator $S^2$ of the experimental random error by using data corresponding the same spectral, spatial and phase points acquired in several recurring experiments. In this estimator not only instrumental noises but also light source possible instability is taken into account.

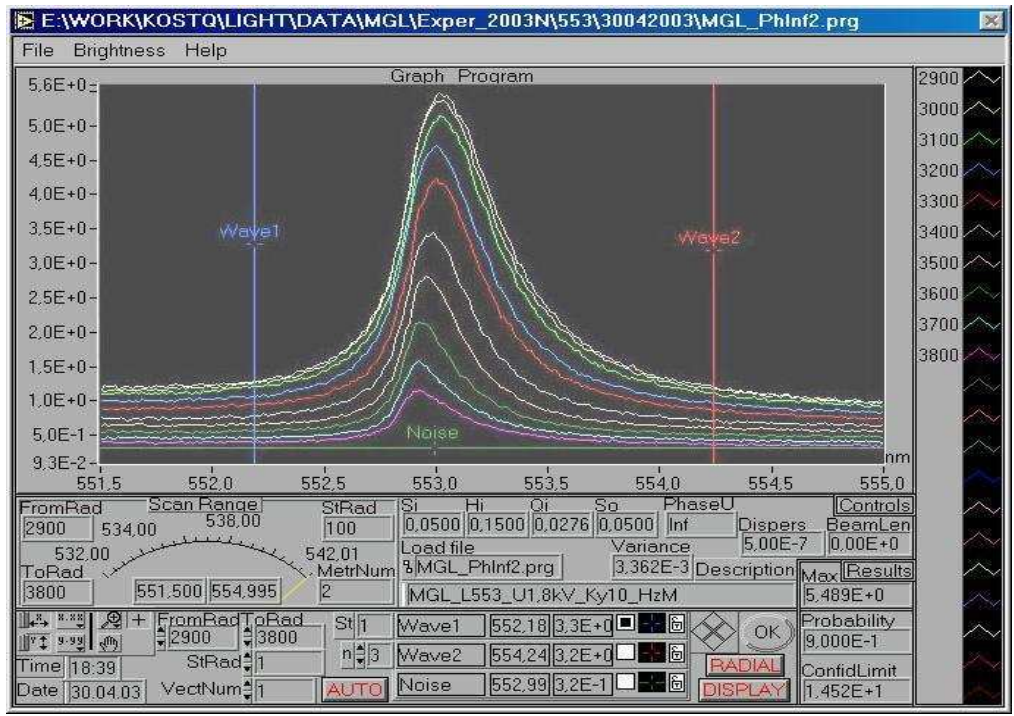

Fig1. VI of data loading for subsequent analysis

**2. Joint data processing algorithm.**

The original method of joint data processing was applied to data arrays containing spatial, spectral and temporal distributions of source surface radiance. (The cylindrical symmetry of the source was supposed). The method is based on the principal component theory [6]. The method takes into account the correlation between spectral emissivities inside spectral line profiles and allows to decrease the processing time as well as the noise influence on the results of instrumental distortion elimination and Abel inversion [7-9].

Let us denote as $F(\lambda_i,x_k)$ the large data arrays of visible spectral intensities, measured along a series of chords, perpendicular to plasma axis ($x_k$ - the displacement of chord from the plasma centerline, $\lambda_i$ - the wavelength of spectral point inside the profile). Instead of instrumental distortion elimination from each profile $F_k(\lambda)$, and then Abel inversion for each wave length $\lambda_i$ we can spread out $F(\lambda_i,x_k)$ on eigenvectors $U_p$ of covariance matrix of array $F(\lambda_i,x_k)$, corresponding to eigenvalues p, superior the estimator $S^2$ of the experimental random error. (As a rule, the number of such vectors q is much less than m - the number of position $x_k$).

$$F_{i,k} = \overline{F}_k + \sum_{p=1}^{q} M_{i,p} U_{p,k}, \qquad (1)$$

Here $\overline{F}_k$ is the spatial visible intensity distribution, average on all spectral intervals, $M_{i,p}$ - factors of decomposition on eigenvectors $U_p$ for various spectral components.

$$M_{i,p} = \sum_{k=1}^{m}(F_{i,k} - \overline{F}_k)U_{p,k} \qquad (2)$$

From these factors the instrumental distortion is to be excluded. The free from distortion factors are designated as $Q_{pi}$. To eliminate instrumental distortion from $M_{i,p}$ we applied the standard method of solving of the integral equation of convolution by Fourier transform [9].

So the required profiles $\varepsilon(\lambda_i,r_j)$ in various plasma points r are:

$$\varepsilon(\lambda_i, r) = R\{\overline{F}\} + \sum_{p=1}^{q} Q_{p\,i}(\lambda)\, R\{U_p\} \qquad (3)$$

Here $R$ is the operator of Abel inversion.

$$R(b) = -\frac{1}{\pi}\int_{r}^{L} \frac{db(x)}{dx}\frac{dx}{\sqrt{(x^2 - r^2)}}, \qquad (4)$$

(where b is $\overline{F}_k$ or $U_p$, L is the radius of discharge tube).

Obviously, in this case the integral Abel equation may be transformed to system of the linear algebraic equations.

The obtained spectral emissivities (3) are used for the determination of plasma parameters in the selected plasma point. (The spatial resolution of the imaging system was 0.02 cm at least).

### 4. The experimental results and discussion.

We investigated a mercury high pressure lamp with addition of thallium iodide which operates at input power of 250 W (50 Hz). The inner tube radius was 0.8 cm. The position of lamp was vertical. All results correspond to the central cross section.

The ground state Hg atom density was determined by the width of mercury spectral lines 577 nm ($6^3D_2$–$6^1P_1$) and 579 nm ($6^1D_2$–$6^1P_1$). The broadening of these lines is the resonance one [7]. So there are no shifts in profiles corresponding to different spatial points and different time moments. There is also no dependence of line width on spatial points and time moments. It is in accordance with our previous work [10], in which it was shown by analysis of plasma interference data that in such sources mercury atoms density is almost constant over the current period and over the most part of the arc, atoms density changes sharply only near the source wall, where there is no radiation. Using this line width we estimated mercury atoms density in the ground state as $N_{Hg}=(6.8\pm0.2)\cdot10^{18}$ cm$^{-3}$.

In accordance to the model from [4] we supposed that mercury atom densities in metastable states $6^3P_2$ and $6^3P_0$ are determined by electron temperature. These levels are lower for spectral lines 546.1 nm and 404.6 nm, which are broadened by reabsorption. For such cases Abel inversion cannot be applied and we used simulation of surface radiance spectral profile for estimation of electron temperature radial distribution [11]. The transition probabilities were obtained from NIST Atomic Spectra Database [12], estimation of the Van-der-Waals broadening constant was supposed to be $2.2\cdot10^{-30}$ cm$^6$/s [13] and ground state densities were determined by resonance broadening. The best accordance between the measured profiles and the calculated ones was obtained when electron temperature (T) radial dependence was supposed as

$$T(r)=T_0-(T_0-T_w)(r/L)^3, \qquad (5)$$

where $T_w=1000$ K is the temperature on the tube wall, L is the wall radius, $T_0$ - is the temperature in the discharge center. $T_0=(5300\pm100)$K at the time moment when the discharge current and light emission are maximal and $T_0=(4100\pm100)$K when they are minimal (Fig.2).

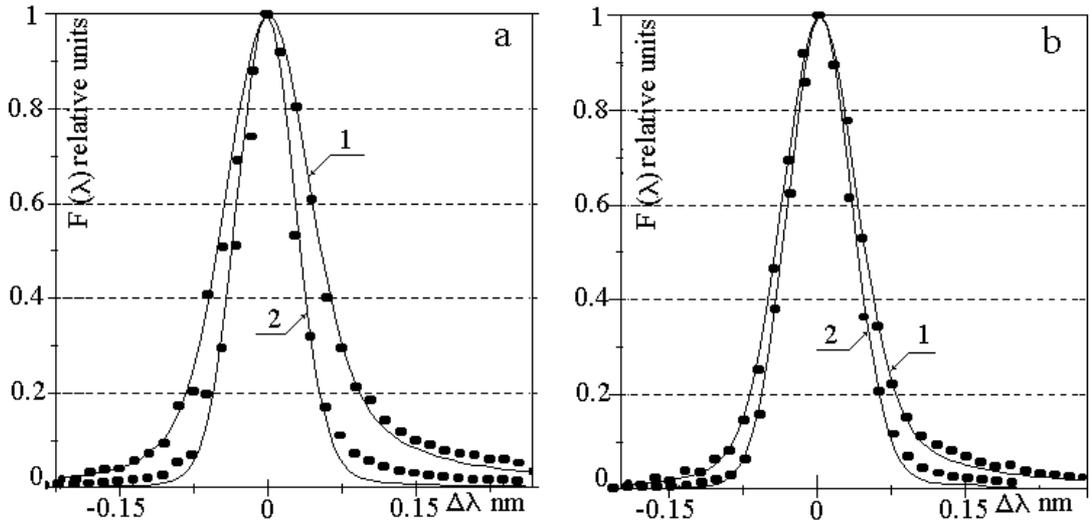

Fig 2. Experimental (points) and calculated (lines) spectral line profiles for maximal (1) and minimal (2) current phases (a - 546.1 nm, b - 404.6 nm)

The high state densities $N_k(r)$ were received from radial and temporal dependencies of integral line intensities.

$$N_k(r)=\frac{4\pi\int\varepsilon(\lambda,r)d\lambda}{A_{k,i}(hc/\lambda_0)}, \qquad (6)$$

where $A_{k,i}$ is transition probability [12] and $hc/\lambda_0$ - is quantum energy for the spectral line. The sharp difference of experimental densities and calculated ones in the frames of LTE model may be explained by errors in transition probabilities or ground state densities, but the time modulation of high state density is independent of these values. In the table 1 the measured densities and their time modulation for the discharge center are presented.

Table 1. The excited state densities for the center of discharge for two current phases. (The random relative error is less than 20%)

| level (line, nm) | E (eV) Excitation energy | N max (см$^{-3}$) | N min (см$^{-3}$) | Nmax/Nmin LTE | Nmax/Nmin Measured |
|---|---|---|---|---|---|
| mercury | | | | | |
| $7^1S_0$ (407.8) | 7.90 | $2.8 \cdot 10^9$ | $1.1 \cdot 10^8$ | 56 | 25 |
| $7^1D_2$ (434.7) | 9.55 | $1.0 \cdot 10^9$ | $3.4 \cdot 10^7$ | 107 | 29 |
| $6^3D_2$ (576.9) | 8.83 | $2.5 \cdot 10^{10}$ | $8.0 \cdot 10^8$ | 87 | 31 |
| $6^1D_2$ (578.9) | 8.82 | $1.0 \cdot 10^{10}$ | $3.0 \cdot 10^8$ | 87 | 33 |
| thallium | | | | | |
| $8^2P_{3/2}$ (654.9) | 5.18 | $4.0 \cdot 10^{11}$ | $4.0 \cdot 10^{10}$ | 40 | 10 |
| $9^2P_{3/2}$ (552.7) | 5.52 | $7.6 \cdot 10^{10}$ | $9.2 \cdot 10^9$ | 51 | 8.3 |

One can see that this high state densities differ from LTE.

To check the Saha balance between electron and high lying excited states densities in accordance to [4] we determined the products of electron ($n_e$) and ion densities from Saha equation

$$\frac{n_e N_a^+}{N_{ak}} = 2 \frac{g_a^+}{g_{ak}} \left(\frac{m\chi T}{2\pi\hbar^2}\right)^{3/2} \exp\left(-\frac{\Delta E_{ki}}{\chi T}\right), \qquad (7)$$

where $N_a^+$, $g_a^+$ - the density and statistical weight of a-kind ions (Tl$^+$ or Hg$^+$), $N_{ak}$, $g_{ak}$ - the density and statistical weight of a-kind atoms in excited state k, $\Delta E_{ki}$ - the ionization energy from the level k, $\chi$ - the Boltzmann constant, m - the electron mass. We determined $n_e$:

$$n_e = N_{hg}^+ + N_{Tl}^+. \qquad (8)$$

Then we have determined electron density by spectral line shifts.

Thallium line 552.8 nm ($9^2P_{3/2}–7^2S_{1/2}$) and 654.9 nm ($8^2P_{3/2}–7^2S_{1/2}$) are broadened mainly by the Stark-effect, it leads to appreciable shifts of profiles in different spatial points and different current phases (Fig.3).

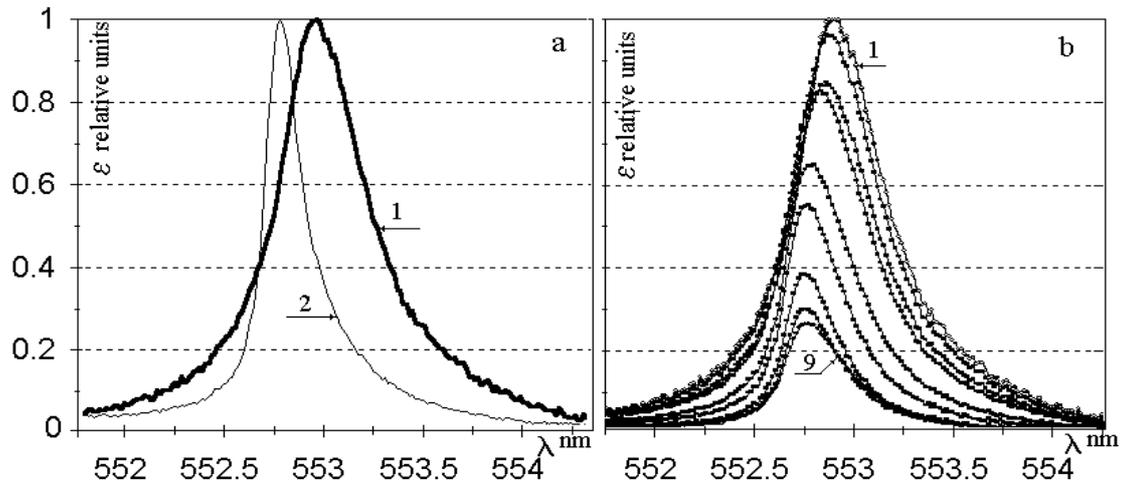

Fig.3. Thallium line 552.8 nm profiles: a) r=0, 1-maximal current, 2- minimal current; b) maximal current, from 1 to 9 - different spatial points from r=0 to r=0.32 cm.

We supposed that the experimental shift of profile maximum Δ is caused by Stark-effect and we estimated the electron densities in the arc cross section by formula [13, 14]:

$$\Delta = 2W n_e \cdot 10^{-17}/1.15, \qquad (9)$$

where Δ - the shift (nm), W - Stark-broadening constant [14], $n_e$ - electron density (cm$^{-3}$). The results for maximal current phase are presented in Fig.4. These results are averaged by two lines and three independent experiments.

In the same figure one can see the electron density, determined from Saha equation.

The product $n_e N_{Tl}^+$ is much more than $n_e N_{hg}^+$ for all Hg high levels, so we estimated $n_e$ as $(n_e N_{Tl}^+)^{1/2}$ and averaged the result by two lines. In the limit of experimental errors for the moment, when discharge current has maximal value, the Saha balance between electron and high lying excited states densities takes place for core part of discharge tube.

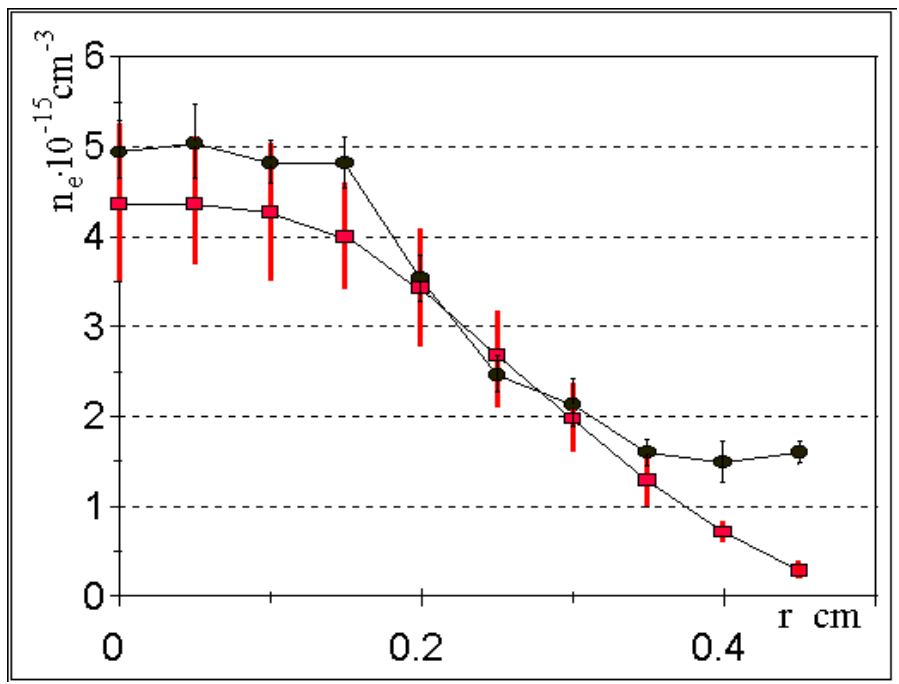

Fig.4. Radial dependence of electron density determined by spectral line shifts (black circles) and Saha equation (red rectangles).

For minimal current phase there is no noticeable spectral line shift except the point r=0 for line 552.8 nm (Δ~0.015 nm, it corresponds $n_e$=4.8·$10^{-14}$ cm$^{-3}$), whereas Saha equation gives for this point $n_e$=(9.2±1.2)·$10^{-14}$ cm$^{-3}$, so we cannot say that Saha balance between electron and high lying excited states densities takes place in minimal current phase.

If we explained the width of these lines only to Stark-broadening we would receive the electron density for r=0 almost three times more than by shift, with increasing of r from r=0 to r=0.24 cm the line width decreases and then begins to increase to the limit of discharge. This effect is noticeable for both current phases. Now we cannot explain this broadening by Van-der-Waals interaction with Hg atoms nor by reabsortion effect (these effects gives less than 10% from the full line width). Perhaps, the effect may be explained by Tl-J interaction, this problem is an object of further investigations.

**Acknowledgments**
Authors wish to acknowledge the support of the Russian Ministry of Science and Technology and the U.S. Civilian Research & Development Foundation for the Independent States of the Former Soviet Union (CRDF) (Award No. PZ-013-02);